%% file: hkim.tex
\documentclass{evn2004}

\usepackage{txfonts}
\usepackage{graphicx}
\begin{document}
   \title{Construction of the Korean VLBI Network (KVN)}

   \author{H.-G. Kim\inst{1}
          \and
          S.-T. Han\inst{1}
          \and
          B. W. Sohn\inst{1,2}
          \and
          S.-J. Oh\inst{1}
          \and
          D.-H. Je\inst{1}
          \and
          S.-O. Wi\inst{1}
          \and
          M.-G. Song\inst{1}
          }

   \institute{
          Korea Astronomy Observatory, 61-1 Hwaam, Yuseong, Daejeon 305-348, Korea
         \and
          Max-Planck-Institut f\"ur Radioastronomie, Auf dem H\"ugel 69, 53121 Bonn, Germany
        }

   \abstract{
Korea's new VLBI project to construct the Korean VLBI Network (KVN) started in 2001, 
as a 7-year project that is fully funded by our government. We plan to build 3 new 
high-precision radio telescopes of 21-m diameter in 3 places in Korea, which will be 
exclusively used for VLBI observations. We will install the 2/8, 22 and 43 GHz HEMT 
receivers within 2007 as a first target, and later we will expand the receiving 
frequency up to 86 and 129 GHz for astronomical, geodetic, and earth science VLBI 
research. The millimeter-wave VLBI will be the ultimate goal of KVN.
For the front-ends, we are going to install a multi-channel receiver system that 
employs low-pass filters within a quasi-optical beam transportation system. This 
receiver system will give reliable phase calibrations for millimeter-wave VLBI as 
well as enable simultaneous multi-frequency band observations. The hard-disk type 
new Mark 5 will be used as the main recorder of KVN. We have completed the design of 
the KVN DAS system of 2 Gsps sampling rate, which will use 4 data streams to meet 
the multi-channel requirement. A VERA type DAS modified for Mark 5 recorder is also 
under consideration. A new correlator project for KVN was recently approved from Korean
government, and will start in the second half of 2004.

}

   \maketitle
%

\section{Introduction}
The first radio astronomical project of Korea Astronomy Observatory (KAO) was the 
construction of the 14-m millimeter wave radio telescope at Daejeon, Korea, which was 
completed about 15 years ago. With the completion of the radome-enclosed 14-m radio 
telescope, the Taeduk Radio Astronomy Observatory (TRAO) was inaugurated as a branch 
of KAO. At present, a dual-channel (100/150 GHz) SIS receiver is being used, which our 
engineering team made, and we are preparing a multi-channel (15-beam) receiver (made 
by FCRAO) to be installed in our 14-m telescope. Our main research subjects are studies 
of interstellar molecular clouds and the physical and chemical processes in dense clouds. 
In particular, the SiO masers, star-forming regions, and interstellar molecular processes 
have been the major focus of our radio astronomy group.

Although we have been managing our millimeter-wave telescope and receiver systems successfully 
so far, our radio group has not yet been involved in any VLBI activities, mainly because of 
our millimeter-wave single-dish facility. Only recently (from 2001) have we started VLBI 
experiments, in millimeter wavelengths, with the Nobeyama VLBI group of Japan with encouraging 
success.

After the completion of 14-m radio telescope project, KAO concentrated its effort mainly 
on the construction of optical telescopes. By the completion of the optical telescope project, 
our radio group submitted a proposal for this Korean VLBI Network (KVN) construction project 
to our government, which was finally approved in 2000.

\section{Scientific Goals of KVN}
The KVN will be the first VLBI facility in Korea, which will be used for VLBI studies in 
astronomy, geodesy, earth science, etc. The general VLBI research targets will also be 
targets for KVN, within the KVN limitations. However, the KVN will be constructed as an 
advanced millimeter-wave (up to $\sim$ 150 GHz) VLBI network. Since millimeter-wave VLBI is 
still in the developing stages around the world, we expect that our KVN will play an important 
role in promoting millimeter-wave VLBI research activities.

Because of the existing 14-m millimeter-wave telescope facility, the research activities 
of the present radio astronomy group of KAO are concentrated on interstellar molecular cloud 
studies. Therefore, one of our main KVN projects will be spectral-line observation focused on 
the interstellar molecular processes at millimeter wavelengths. At the same time, we are 
strengthening our VLBI group by increasing its members in high-energy astrophysics and 
AGN-related research fields, in order to promote research activities in various VLBI subjects.

In geodetic studies, the geodetic society groups in Korea will lead most of the research work. 
Since the tectonic movement of the Korean peninsula, including several important fault plane 
movements, has never been measured, the monitoring of these movements will be an important 
national project, which will be carried out with KVN. KVN will also participate actively 
in international campaigns for geodetic measurements.

\section{Outline of the Project}
In 2001, KVN started a 7-year project to construct the first VLBI facilities in Korea, 
which include 3 new radio telescopes and VLBI receiving systems to be dedicated exclusively 
to VLBI observations.

\subsection{KVN Sites}
The three KVN observatory sites, Yonsei University at Seoul, University of Ulsan at Ulsan, and 
Tamna University at Jeju, have been selected from among the many universities and institutes 
who wanted to invite our telescopes. Fig. 1 shows the location of these three KVN sites. 
The maximum baseline length is about 450 km in the north-south direction, and the site 
parameters are summarized in Table 1 and 2. The site preparation work has already started 
at all three places. In Fig. 2, we show an example of the UV coverage and the synthesized 
beam shape for the three KVN antennas, including the existing 14-m radio telescope at TRAO, 
Daejeon. In future, these 4 antennas will be connected using the optical fiber information 
networks in Korea, partly for e-VLBI, and partly for the real-time operations of KVN.
\begin{figure}
   \centering
   \includegraphics[width=.5\textwidth]{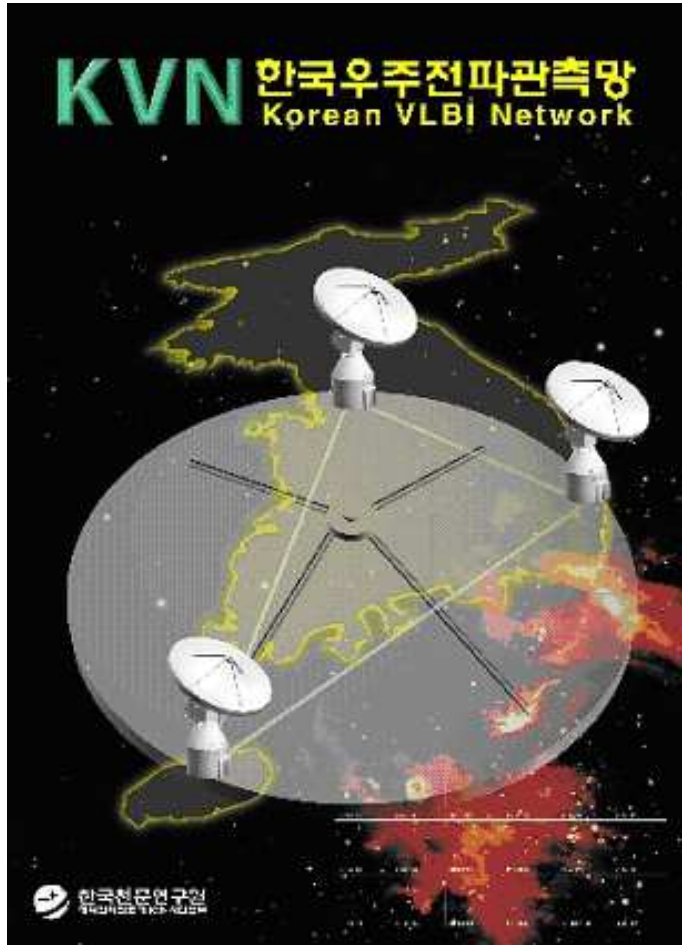}
      \caption{Locations of KVN observatories}
         \label{fig1}
\end{figure}

\begin{table}
\caption{Site locations of KVN and TRAO}
\begin{center}
\renewcommand{\arraystretch}{1.4}
\setlength\tabcolsep{5pt}
\begin{tabular}{lrr}
\hline\noalign{\smallskip}
KVN & Longitude & Latitude \\
Site & $^{\circ}$~'~"~E & $^{\circ}$~'~"~N \\
\hline\noalign{\smallskip}
Yonsei (Seoul) & 126~ 56~ 35 & 37~ 33~ 44 \\
Ulsan (Ulsan) & 129~ 15~ 04 & 35~ 32~ 33 \\
Tamna (Jeju) & 126~ 27~ 43 & 33~ 17~ 18 \\
TRAO (Daejeon) & 127~ 22~ 19 & 36~ 23~ 53 \\
\hline
\end{tabular}
\end{center}
\label{Tab1}
\end{table}
\begin{table}
\caption{Baselines of KVN observatories and TRAO}
\begin{center}
\renewcommand{\arraystretch}{1.4}
\setlength\tabcolsep{5pt}
\begin{tabular}{lllll}
\hline\noalign{\smallskip}
KVN & Baseline \\
Observatory & Yonsei & Ulsan & Tamna & TRAO \\
\hline\noalign{\smallskip}
Yonsei (Seoul) & -- & 305.2 & 477.7 & 135.1 \\
Ulsan (Ulsan) & 305.2 & -- & 358.5 & 194.2 \\
Tamna (Jeju) & 477.7 & 358.5 & -- & 356.0 \\
TRAO (Daejeon) & 135.1 & 194.2 & 356.0 & -- \\
\hline
\end{tabular}
\end{center}
\label{Tab2}
\end{table}
%
\begin{figure}
   \centering
   \includegraphics[width=.5\textwidth]{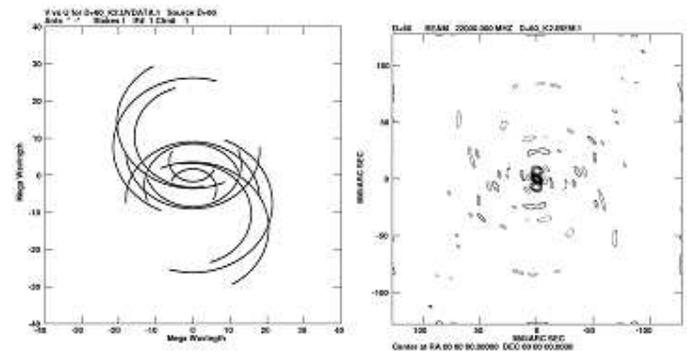}
      \caption{Example UV coverage and synthesized beam shape 
for a source at decl.= ${\rm 60^{\circ}}$ for KVN and TRAO}
         \label{fig2}
\end{figure}

\subsection{Antennas}
We plan to build 3 new high-precision Cassegrain type radio telescopes each of 21-m diameter, 
having reasonable efficiencies at frequencies above 100 GHz (total rms ${\rm \le 150 ~\mu m }$). 
Fig. 3 shows bird's-eye view of KVN antenna.

The maximum slewing speed is 3${\rm ^{\circ}/sec}$ in both AZ and EL with an acceleration of 
about 3${\rm ^{\circ}/sec^2}$. 
The requested wind speed tolerance is about 10~(20) m/s for precision (limited) observations, 
and 90 m/s for survival. We plan to start the foundation construction in 2004, and the antenna 
installations in 2005.

\begin{figure}
   \centering
   \includegraphics[width=.5\textwidth]{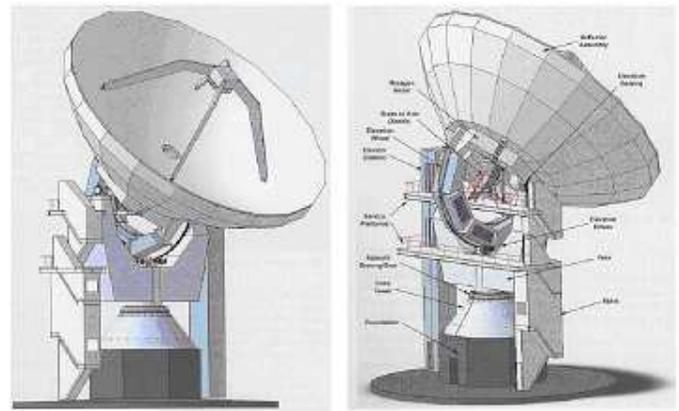}
      \caption{Bird's-eye view of KVN antenna}
         \label{fig3}
\end{figure}

\subsection{RF Receivers}
For the KVN front-ends, several cryogenic HEMT receivers will be installed at the Cassegrain 
focus for 2/8, 22, 43, 86 and 129 GHz operations. The 2/8 GHz receivers will be mainly for 
geodetic observations. The 22 and 43 GHz receivers will be installed first, to set up the 
antennas and for the initial VLBI observations. Since our main goal will be mm-VLBI, the 86 
and 129 GHz receivers will be installed as soon as the full test of the KVN system is 
completed. Some parameters of the first receivers are summarized in Table 3.

\begin{table}
\caption{KVN receiver specifications}
\begin{center}
\renewcommand{\arraystretch}{1.4}
\setlength\tabcolsep{5pt}
\begin{tabular}{lcccc}
\hline\noalign{\smallskip}
Freq Band & S Band & X Band & K Band & Q Band\\
\hline\noalign{\smallskip}
Freq [GHz] & 2.2 - 2.8 & 8 - 9 & 21.5 - 23.5 & 42 - 44\\
Rx Noise~(K)& $<$ 25 & $<$ 25 & $<$ 30 & $<$ 50\\
1st IF~(GHz)& 2.5 & 8.5 & 8.5 & 8.5\\
IF BW~(GHz)& 0.6 & 1 & 2 & 2\\
IF Power~(dBm)& -25 & -25 & -25 & -25\\
Pol.~(CP) & full & full & full & full\\
\hline
\end{tabular}
\end{center}
\label{Tab3}
\end{table}

For KVN, which is designed for mm-VLBI, we plan to adopt a multi-channel quasi-optical beam 
transportation system which can be used for phase calibration in millimeter and 
sub-millimeter-wave VLBI, without losing observing time, and without the necessity to look 
for reference sources. In addition, this method enables us to observe several frequencies 
simultaneously.

Fig. 3 shows the conceptual design of the KVN beam transporting system employing 
frequency-selective surfaces - low-pass filters (LPFs) (Goldsmith 1998).

The lowest 22 GHz band can be used as a phase calibration reference for the higher frequency 
band observations, made toward the same source at the same time. Although there are some 
limitations (Sasao 2003), this multi-channel idea would give very reliable phase corrections 
in mm-VLBI.
\begin{figure}
   \centering
   \includegraphics[width=.5\textwidth]{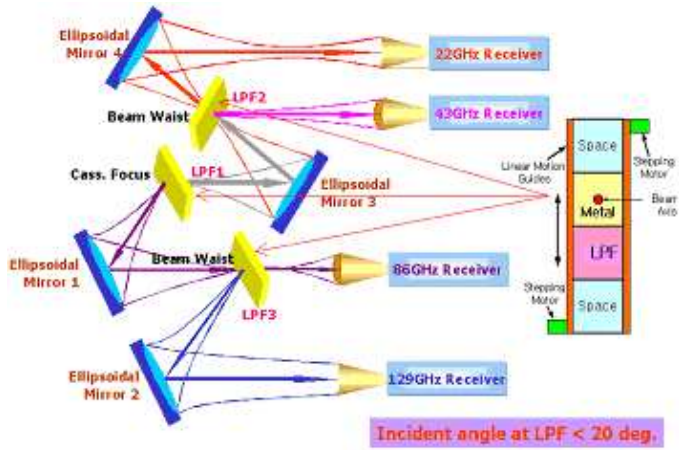}
      \caption{Conceptual design of KVN quasi-optics}
         \label{fig4}
   \end{figure}

\subsection{KVN DAS and Recorder}
We are developing the KVN data acquisition system (KVN DAS). Fig. 5 shows the configuration 
of KVN DAS. For our multi-channel receiver system, we employ four high-speed samplers operated 
at 1 Gsps. These four data streams of 2 Gbps will be transported via optical fibers to the 
operation building, and then distributed among sixteen FIR digital filters. With these filters, 
we can choose a passband whose center frequency is arbitrarily programmable in the input 
bandwidth, and then resample the filtered data at 2 bits per sample. These resampled data 
streams are then formatted and sent to the recorder. We will also prepare the digital 
spectrometer for single-dish operation and total-power measurements.

As for the KVN recorder, we plan to use the new Mark5 recorder (Whitney 2003). KVN is involved 
in the development consortium for the Mark5, led by Haystack Observatory of MIT. The 
characteristics of the Mark5, which has a 1 Gbps rate using a standard PC and 16 hard disks 
with a VSI interface, are well explained in the Haystack web site. We plan to reproduce the 
Mark5 for KVN.

We have just started the design of the new KVN correlator, this project is in the very first 
planning stages, at present.
\begin{figure}
   \centering
   \includegraphics[width=.5\textwidth]{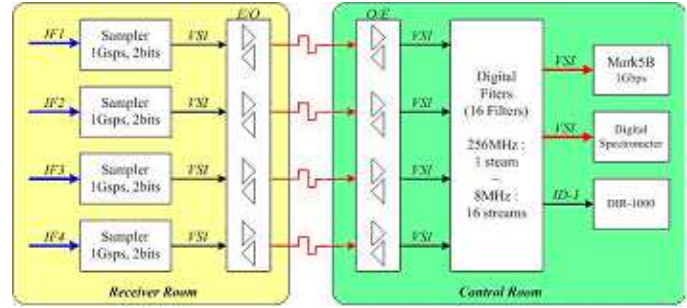}
      \caption{Schematic diagram of KVN DAS}
         \label{fig5}
   \end{figure}

\section{Summary}
The construction of KVN is underway as scheduled. The KVN is the first VLBI facility in Korea, 
which will be used for astronomy, geodesy, and earth science. It is our wish that KVN will be 
one of the best VLBI systems in the world, and we will actively participate in various VLBI 
programs after completion. We hope that KVN will play a central role in promoting the VLBI 
research work of Korean societies and international collaborations, through the KVN Research 
Center that will be built in Seoul. Various support and manpower exchange programs to stimulate 
VLBI activities will be organized by the KVN project. Finally, for the success of the KVN 
project, it must be essential for us to collaborate with, and get many suggestions and supports 
from, the leading institutes in VLBI research around the world.

%

\cleardoublepage

\end{document}